\documentclass[12pt,openbib]{article}%
\usepackage{amsmath}
\usepackage{amsfonts}
\usepackage{amssymb}
\usepackage{graphicx}
\usepackage{natbib}
\usepackage{setspace}%
\newtheorem{theorem}{Theorem}

\newenvironment{proof}[1][Proof]{\noindent\textbf{#1.} }{\ \rule{0.5em}{0.5em}}

\newcommand{\jmdouble}{\doublespacing}
\renewcommand{\jmdouble}{\relax}
\jmdouble
\begin{document}

\title{On probabilities for separating sets of order statistics\footnotemark[1]}
\author{D. H. Glueck\footnotemark[2]
\and A. Karimpour-Fard\footnotemark[2]
\and J. Mandel\footnotemark[2]
\and K. E. Muller\footnotemark[3]}
\date{June 24, 2007}
\maketitle

\footnotetext[1]{Deborah H. Glueck is Assistant Professor, Department of
Preventive Medicine and Biometrics, University of Colorado at Denver and
Health Sciences Center, Campus Box B119, 4200 East Ninth Avenue, Denver,
Colorado 80262 (e-mail: Deborah.Glueck@uchsc.edu). Anis Karimpour-Fard is a
graduate student in Bioinformatics, Department of Preventive Medicine and
Biometrics, University of Colorado at Denver and Health Sciences Center,
Campus Box B119, 4200 East Ninth Avenue, Denver, Colorado 80262 (e-mail: Anis
Karimpour-Fard@uchsc.edu). Jan Mandel is Professor, Department of Mathematics,
Adjunct Professor, Department of Computer Science, and Director of the Center
for Computational Mathematics, University of Colorado at Denver and Health
Sciences Center, Campus Box 170, Denver, Colorado 80217-3364
(e-mail:Jan.Mandel@cudenver.edu). Keith E. Muller is Professor and Director of
the Division of Biostatistics, Department of Epidemiology and Health Policy
Research, University of Florida, 1329 SW 16th Street Room 5125, PO Box 100177
Gainesville, FL 32610-0177 (e-mail:Keith.Muller@biostat.ufl.edu)
\par
Glueck was supported by NCI K07CA88811. Mandel was supported by NSF-CMS
0325314. Muller was supported by NCI P01 CA47 982-04, NCI R01 CA095749-01A1
and NIAID 9P30 AI 50410.
\par
The authors thank Professor Gary Grunwald for his helpful comments.}

\footnotetext[2]{University of Colorado at Denver and Health Sciences Center}
\footnotetext[3]{University of Florida}


\begin{abstract}
Consider a set of order statistics that arise from sorting samples from two
different populations, each with their own, possibly different distribution
function. \ The probability that these order statistics fall in disjoint,
ordered intervals, and that of the smallest statistics, a certain number come
from the first populations, are given in terms of the two distribution
functions. The result is applied to computing the joint probability of the
number of rejections and the number of false rejections for the
Benjamini-Hochberg false discovery rate procedure.

\end{abstract}



\section{Introduction}

Glueck et al.\ (2006b) gave explicit expressions for the probability that
arbitrary subsets of order statistics fall in disjoint, ordered intervals on
the set of real numbers. \ In this paper, we extend this work and consider two
sets of real valued, independent but not necessarily identically distributed
random variables. \ We give expressions in terms of cumulative distribution
functions for the probability that arbitrary subsets of order statistics fall
in disjoint, ordered intervals, and that of the smallest statistics, a certain
number come from one set. We have been unable to find any previous papers on
this topic. \ This problem is of interest in calculating probabilities for the
Benjamini and Hochberg (1995) multiple comparisons procedure. \label{fall}

\section{A simple example}

\label{sec:example} Consider the following simple example. Let\ $X_{1}$,
$X_{2}$ $\in\left[  0,1\right]  $ be independent random variables. Denote by
$F_{X_{1}}\left(  x_{1}\right)  $ and $F_{X_{2}}\left(  x_{2}\right)  $ the
marginal cumulative distribution functions and by $F_{X_{1},X_{2}}\left(
x_{1},x_{2}\right)  $ the joint cumulative distribution function of $X_{1}$
and $X_{2}$. \ Assume that the cumulative distribution functions are
continuous. \ Let $Y_{1}=\min\left\{  X_{1},X_{2}\right\}  $ and let
$Y_{2}=\max\left\{  X_{1},X_{2}\right\}  $ be the order statistics. \ For
$i=1,2$, write the marginal cumulative distribution function of $Y_{i}$ as
$F_{Y_{i}}\left(  y_{i}\right)  $, and the joint cumulative distribution
function as $F_{Y_{1},Y_{2}}\left(  y_{1},y_{2}\right)  $, for $y_{1}\leq
y_{2}$. \ This joint cumulative distribution function is also continuous
(David, 1981, p. 10).

Choose numbers $b_{1}<b_{2}$, $b_{1}$, $b_{2}\in\left(  0,1\right)  $. We wish
to find the probabilities
\begin{align}
\mathcal{A}  &  =\Pr\left\{  \left(  y_{1}<b_{1}\right)  \wedge\left(
y_{2}>b_{2}\right)  \right\}  ,\\
\beta &  =\Pr\left\{  \left(  y_{1}<b_{1}\right)  \wedge\left(  y_{2}%
>b_{2}\right)  \wedge\left(  x_{1}<b_{1}\right)  \right\}
\end{align}
and
\begin{equation}
\gamma=\Pr\left\{  \left(  y_{1}<b_{1}\right)  \wedge\left(  y_{2}%
>b_{2}\right)  \wedge\lnot\left(  x_{1}<b_{1}\right)  \right\}  .
\end{equation}
and express them in terms of the distribution functions $F_{X_{1}}$ and
$F_{X_{2}}$. First, we will find the probabilities directly. So,%
\begin{align}
\beta &  =\Pr\left\{  \left(  x_{1}<b_{1}\right)  \wedge\left(  x_{2}%
>b_{2}\right)  \right\} \nonumber\\
&  =F_{X_{1}}\left(  b_{1}\right)  \left[  1-F_{X_{2}}\left(  b_{2}\right)
\right]  \label{eq:x}%
\end{align}
and
\begin{align}
\gamma &  =\Pr\left\{  \left(  x_{1}>b_{2}\right)  \wedge\left(  x_{2}%
<b_{1}\right)  \right\} \nonumber\\
&  =\left[  1-F_{X_{1}}\left(  b_{2}\right)  \right]  F_{X_{2}}\left(
b_{1}\right)  . \label{breakfast}%
\end{align}

Equations (\ref{eq:x}) and (\ref{breakfast}) follow directly from the
independence of the random variables, and the definition of the cumulative
distribution functions. Since \
\begin{align}
&  \left\{  \left(  y_{1}<b_{1}\right)  \wedge\left(  y_{2}>b_{2}\right)
\right\}  =\label{eq:union}\\
&  \qquad\left\{  \left(  y_{1}<b_{1}\right)  \wedge\left(  y_{2}%
>b_{2}\right)  \wedge\left(  x_{1}<b_{1}\right)  \right\}  \cup\left\{
\left(  y_{1}<b_{1}\right)  \wedge\left(  y_{2}>b_{2}\right)  \wedge
\lnot\left(  x_{1}<b_{1}\right)  \right\} \nonumber
\end{align}
and the union is disjoint, it follows that
\begin{equation}
\mathcal{A}=\beta+\gamma. \label{eq:sum}%
\end{equation}

For a problem with more than two order statistics, the number of cases one
needs to consider and the number of possible combinations of statistics,
subsets, and bounds makes a direct approach impractical. An algorithmic
approach to obtaining $\gamma$ and $\beta$ will allow the generalization to an
arbitrary number of order statistics.

Using the assumption that the distribution functions are continuous, simple
set operations, and the definition of distribution function, we obtain that
the probability of the union (\ref{eq:union}) is%
\begin{align}
\mathcal{A}  &  =\Pr\left\{  \left(  y_{1}<b_{1}\right)  \wedge\lnot\left(
y_{2}<b_{2}\right)  \right\} \\
&  =\Pr\left\{  y_{1}<b_{1}\right\}  -\Pr\left\{  \left(  y_{1}<b_{1}\right)
\wedge\left(  y_{2}<b_{2}\right)  \right\} \\
&  =F_{Y_{1}}\left(  b_{1}\right)  -F_{Y_{1},Y_{2}}\left(  b_{1},b_{2}\right)
. \label{eq:Y}%
\end{align}
\ The cumulative distributions of the order statistics can be written (Bapat
and Beg, 1989), \ \ %

\begin{equation}
F_{Y_{1}}\left(  b_{1}\right)  =F_{X_{1}}\left(  b_{1}\right)  \left[
1-F_{X_{2}}\left(  b_{1}\right)  \right]  +\left[  1-F_{X_{1}}\left(
b_{1}\right)  \right]  F_{X_{2}}\left(  b_{1}\right)  \label{red}%
\end{equation}%
\begin{equation}
F_{Y_{1},Y_{2}}\left(  b_{1},b_{2}\right)  =F_{X_{1}}\left(  b_{1}\right)
\left[  F_{X_{2}}\left(  b_{2}\right)  +F_{X_{2}}\left(  b_{1}\right)
\right]  -\left[  F_{X_{1}}\left(  b_{2}\right)  -F_{X_{1}}\left(
b_{1}\right)  \right]  F_{X_{2}}\left(  b_{1}\right)  . \label{cucumber}%
\end{equation}
Then, substituting Equations (\ref{red}) and (\ref{cucumber}) into Equation
(\ref{eq:Y}), we can write $\mathcal{A}$ in terms of the distribution
functions of $X_{1}$ and $X_{2}$,
\begin{align}
\mathcal{A}=  &  F_{X_{1}}\left(  b_{1}\right)  \left[  1-F_{X_{2}}\left(
b_{1}\right)  \right]  +\left[  1-F_{X_{1}}\left(  b_{1}\right)  \right]
F_{X_{2}}\left(  b_{1}\right) \\
&  -F_{X_{1}}\left(  b_{1}\right)  \left[  F_{X_{2}}\left(  b_{2}\right)
-F_{X_{2}}\left(  b_{1}\right)  \right]  -\left[  F_{X_{1}}\left(
b_{2}\right)  -F_{X_{1}}\left(  b_{1}\right)  \right]  F_{X_{2}}\left(
b_{1}\right) \nonumber\\
=  &  F_{X_{1}}\left(  b_{1}\right)  \left[  1-F_{X_{2}}\left(  b_{2}\right)
\right]  +\left[  1-F_{X_{1}}\left(  b_{2}\right)  \right]  F_{X_{2}}\left(
b_{1}\right)  \text{ .} \label{eq:A-sum}%
\end{align}

We now interpret the terms in the sum in Equation (\ref{eq:A-sum}). The term
that includes $F_{X_{1}}\left(  b_{1}\right)  $ as a factor is the probability
of an event in which $x_{1}<b_{1}$ occurs, and the term that includes
$1-F_{X_{1}}\left(  b_{2}\right)  $ as a factor is the probability of an event
in which $x_{1}>b_{2}$. Since $b_{1}<b_{2}$, the two events are disjoint, and,
consequently, (\ref{eq:sum}) follows again.

To summarize, we have expressed the probability in terms of the joint
distribution of the order statistics, which was in turn written in terms of
the distribution functions of the random variables. Finally, by recognizing
terms that corresponded to a partition, we decomposed $\mathcal{A}$ into a sum
of $\beta$ and $\gamma$, the two probabilities of interest.

\section{General case}

\label{sec:general}The logic used in this simple, two random variables example
can be generalized to an arbitrary number of random variables. \ Consider a
set of order statistics that arise from sorting samples from two different
populations, each with their own, possibly different distribution function.
\ We wish to find the probability that these order statistics fall in a given
union of intervals, and that of the smallest statistics, a certain number come
from one population.

For this general case, we need to introduce some notation and definitions.
\ Let $X_{i}$, $i=1,\ldots m$, be independent but not necessarily identically
distributed real valued random variables with values in the interval $[0,1]$
and continuous cumulative distribution functions $F_{X_{i}}\left(
x_{i}\right)  $. Partition the set $\left\{  X_{1},X_{2},\ldots,X_{m}\right\}
$ into two subsets,
\begin{equation}
S_{1}=\left\{  X_{1},X_{2},\ldots,X_{n}\right\}  ,\quad S_{2}=\left\{
X_{n+1},X_{n+2},\ldots,X_{m}\right\}  .
\end{equation}
For example, one can consider measurements for males or females, or for two
different populations of breast cancer, slow or fast growing. \ The order
statistics $Y_{1},Y_{2},\ldots,Y_{m}$ are random variables defined by sorting
the values of $X_{i}$. Thus $Y_{1}\leq Y_{2}\leq\ldots\leq Y_{m}$. Denote the
realizations of the order statistics by $y_{1}\leq y_{2}\leq\ldots\leq y_{m}$.

The arguments of the joint cumulative distribution function of order
statistics are customarily written omitting redundant arguments; thus for
$1\leq e\leq m$ let $1\leq n_{1}<n_{2}<\cdots<n_{_{e}}\leq m$, denote the
indices of the order statistics of interest. The joint cumulative distribution
function of the set $\left\{  Y_{n_{1}},Y_{n_{2}},\ldots,Y_{n_{e}}\right\}  $,
which is a subset of the complete set of order statistics, is defined as%
\begin{equation}
F_{Y_{n_{1}},\ldots Y_{n_{e}}}\left(  y_{1},\ldots,y_{e}\right)  =\Pr\left(
\left\{  Y_{n_{1}}\leq y_{1}\right\}  \cap\left\{  Y_{n_{2}}\leq
y_{2}\right\}  \cap\cdots\cap\left\{  Y_{n_{e}}\leq y_{e}\right\}  \right)  .
\end{equation}

Suppose we are given $s\leq m$ disjoint intervals
\begin{equation}
\left(  c_{q},d_{q}\right)  ,\ \quad0=c_{1}<d_{1}<c_{2}<\cdots<c_{s}<d_{s}=1,
\end{equation}
and integers
\begin{equation}
k_{q}\geq0,\quad\sum_{q=1}^{s}k_{q}=m,
\end{equation}
where $k_{0}=0$ and $k_{q}$ is the number of order statistics that fall in the
$q^{th}$ interval. Define $w_{q,1}=1+\sum_{i=1}^{q-1}k_{i}$, and $w_{q,k_{q}%
}=\sum_{i=1}^{q}k_{i}$ to be the subscripts of the largest and smallest order
statistics, respectively, that fall in the $q^{\mathrm{th}}$ interval. In the
case when $k_{q}=1$, we have $w_{q,1}=w_{q,k_{q}}$. Using this notation, the
event that exactly $k_{q}$ of the order statistics fall in the $q^{th}$
interval is
\begin{equation}
\left\{  c_{1}<Y_{w_{1,1}}<\cdots<Y_{w_{1,k_{1}}}<d_{1}\wedge\cdots\wedge
c_{s}<Y_{w_{s,1}}<\cdots<Y_{w_{s,k_{s}}}<d_{s}\right\}  , \label{eq:def-int}%
\end{equation}
or, in a more compact notation (\ref{eq:E}) below. Now let $B$ be another
random event. The following theorem gives the probability of this event
intersected with the event (\ref{eq:def-int}), in terms of the cumulative
distribution functions of the order statistics relative to the event $B$. This
distrubution function is defined by
\begin{align}
&  F_{Y_{n_{1}},\ldots Y_{n_{e}};B}\left(  y_{1},\ldots,y_{e}\right)
\label{eq:cumulative}\\
&  \qquad=\Pr\left(  \left\{  Y_{n_{1}}\leq y_{1}\right\}  \cap\left\{
Y_{n_{2}}\leq y_{2}\right\}  \cap\cdots\cap\left\{  Y_{n_{e}}\leq
y_{e}\right\}  \cap B\right) \nonumber
\end{align}
Contrary to the usual convention, we do not require that the indices of the
order statistics in the cumulative distribution function (\ref{eq:cumulative}%
)\ are sorted, because that would result in a complication of the notation in
the next theorem (additional renumbering of the arguments).

\begin{theorem}
\label{theorem:1}Denote the event%
\begin{equation}
E=\bigcap\limits_{q=1}^{s}\left(  \left\{  c_{q}<Y_{w_{q,1}}\right\}
\cap\bigl\{Y_{w_{q,k_{q}}}<d_{q}\bigr\}\right)  . \label{eq:E}%
\end{equation}
Then
\begin{align}
\qquad\Pr\left(  E\cap B\right)  =  &  F_{Y_{w_{1,k_{1}}},Y_{w_{2,k_{2}}%
},\ldots,Y_{w_{s,k_{s}}};B}\left(  d_{1},d_{2},\ldots,d_{q}\right) \\
&  -\sum_{i=1}^{s}F_{Y_{w_{1,k_{1}}},Y_{w_{2,k_{2}}},\ldots,Y_{w_{s,k_{s}}%
},Y_{w_{i,k_{i}}};B}\left(  d_{1},d_{2},\ldots,d_{q},c_{q}\right) \nonumber\\
&  +\sum\limits_{\substack{r,t=1\\r<t}}^{s}F_{Y_{w_{1,k_{1}}},Y_{w_{2,k_{2}}%
},\ldots,Y_{w_{s,k_{s}}},Y_{w_{r,1}},Y_{w_{t,1}};B}\left(  d_{1},d_{2}%
,\ldots,d_{q},c_{r},c_{t}\right) \nonumber\\
&  \vdots\nonumber\\
&  +\left(  -1\right)  ^{s}F_{Y_{w_{1,1}},Y_{w_{1,k_{1}}},Y_{w_{2,1}%
},Y_{w_{2,k_{2}}},\ldots,Y_{w_{s,1}},Y_{w_{s,k_{s}}};B}\left(  c_{1}%
,d_{1},c_{2},d_{2},\ldots,c_{s},d_{s}\right)  .\nonumber
\end{align}

\end{theorem}

\begin{proof}
By standard set operations,%
\begin{equation}
E=\bigcap\limits_{q=1}^{s}\left\{  c_{q}<Y_{w_{q,1}}\right\}  \cap
\bigcap\limits_{q=1}^{s}\left\{  Y_{w_{q,k_{q}}}<d_{q}\right\}
\end{equation}
and
\begin{equation}
\bigcap\limits_{q=1}^{s}\left\{  c_{q}<Y_{w_{q,1}}\right\}  =\bigcap
\limits_{q=1}^{s}\left\{  Y_{w_{q,1}}\leq c_{q}\right\}  ^{C}=\left(
\bigcup\limits_{q=1}^{s}\left\{  Y_{w_{q,1}}\leq c_{q}\right\}  \right)  ^{C},
\end{equation}
where $^{C}$ denotes the complement. Therefore,%
\begin{equation}
E\cap B=\left(  \bigcup\limits_{q=1}^{s}\left\{  Y_{w_{q,1}}\leq
c_{q}\right\}  \right)  ^{C}\cap F, \label{eq:EB}%
\end{equation}
where the event $F$ is defined by
\begin{equation}
F=\bigcap\limits_{q=1}^{s}\left\{  Y_{w_{q,k_{q}}}<d_{q}\right\}  \cap B.
\label{eq:F}%
\end{equation}
By the additivity of probability, it follows from (\ref{eq:EB}) that%
\begin{equation}
\Pr\left(  E\cap B\right)  =\Pr\left(  F\right)  -\Pr\left(  \bigcup
\limits_{q=1}^{s}\left\{  Y_{w_{q,1}}\leq c_{q}\right\}  \cap F\right)
=\Pr\left(  F\right)  -\Pr\left(  \bigcup\limits_{q=1}^{s}A_{q}\right)  ,
\end{equation}
where $A_{q}=\left\{  Y_{w_{q,1}}\leq c_{q}\right\}  \cap F$. Using the
additivity of probability again, we have%
\begin{align}
\Pr\left(  \bigcup\limits_{q=1}^{s}A_{q}\right)   &  =\sum_{q=1}^{s}\Pr\left(
A_{q}\right)  -\sum\limits_{\substack{r,t=1\\r<t}}^{s}\Pr\left(  A_{r}\cap
A_{t}\right)  +\cdots\\
&  +\left(  -1\right)  ^{s}\Pr\left(  \bigcap\limits_{q=1}^{s}A_{q}\right)
\label{eq:overlap}%
\end{align}

Now putting (\ref{eq:F}) -- (\ref{eq:overlap}) together and using the
continuity of the cumulative distribution functions, we obtain%
\begin{align*}
\Pr\left(  E\cap B\right)   &  =\underbrace{\Pr\left(  \bigcap\limits_{q=1}%
^{s}\left\{  Y_{w_{q,k_{q}}}\leq d_{q}\right\}  \cap B\right)  }%
_{F_{Y_{w_{1,k_{1}}},Y_{w_{2,k_{2}}},\ldots,Y_{w_{s,k_{s}}};B}\left(
d_{1},d_{2},\ldots,d_{q}\right)  }\\
&  -\sum_{r=1}^{s}\underbrace{\Pr\left(  \left\{  Y_{w_{r,1}}\leq
c_{r}\right\}  \cap\bigcap\limits_{q=1}^{s}\left\{  Y_{w_{q,k_{q}}}\leq
d_{q}\right\}  \cap B\right)  }_{F_{Y_{w_{1,k_{1}}},Y_{w_{2,k_{2}}}%
,\ldots,Y_{w_{s,k_{s}}},Y_{w_{r,1}};B}\left(  d_{1},d_{2},\ldots,d_{q}%
,c_{r}\right)  }\\
&  +\sum\limits_{\substack{r,t=1\\r<t}}^{s}\underbrace{\Pr\left(  \left\{
Y_{w_{r,1}}\leq c_{t}\right\}  \cap\left\{  Y_{w_{t,1}}\leq c_{t}\right\}
\cap\bigcap\limits_{q=1}^{s}\left\{  Y_{w_{q,k_{q}}}\leq d_{q}\right\}  \cap
B\right)  }_{F_{Y_{w_{1,k_{1}}},Y_{w_{2,k_{2}}},\ldots,Y_{w_{s,k_{s}}%
},Y_{w_{r,1}},Y_{w_{t,1}};B}\left(  c_{1}d_{1},d_{2},\ldots,d_{q},c_{r}%
,c_{t}\right)  }\\
&  \vdots\\
&  +\left(  -1\right)  ^{s}\underbrace{\Pr\left(  \bigcap\limits_{q=1}%
^{s}\left\{  Y_{w_{q,1}}\leq c_{q}\right\}  \cap\bigcap\limits_{q=1}%
^{s}\left\{  Y_{w_{q,k_{q}}}\leq d_{q}\right\}  \cap B\right)  }%
_{F_{Y_{w_{1,1}},Y_{w_{1,k_{1}}},Y_{w_{2,1}},Y_{w_{2,k_{2}}},\ldots
,Y_{w_{s,1}},Y_{w_{s,k_{s}}};B}\left(  c_{1},d_{1},c_{2},d_{2},\ldots
,c_{s},d_{s}\right)  },
\end{align*}
which concludes the proof.

\begin{table}[ptb]
\begin{center}%
\begin{tabular}
[t]{c|c|c|c}%
\multicolumn{1}{c}{} & \multicolumn{1}{c}{$<d_{1}$} & \multicolumn{1}{c}{$\geq
d_{1}$} & \multicolumn{1}{c}{}\\\cline{2-3}%
$S_{1}$ & $j$ & $n-j$ & $n$\\\cline{2-3}%
$S_{2}$ & $k_{1}-j$ & $\left(  m-n\right)  -\left(  k_{1}-j\right)  $ &
$\left(  m-n\right)  $\\\cline{2-3}%
\multicolumn{1}{c}{Total} & \multicolumn{1}{c}{$k_{1}$} &
\multicolumn{1}{c}{$m-k_{1}$} & \multicolumn{1}{c}{$m$}%
\end{tabular}
\end{center}
\caption{Numbers of order statistics from the sets $S_{1}$ and $S_{2}$ in the
interval $\left(  0,d_{1}\right)  $ and outside the interval $\left(
0,d_{1}\right)  $, in the event $B$.}%
\label{tab:decision}%
\end{table}
\end{proof}

From now on assume that $B$ is the event that exactly $j$ elements of $S_{1}$
fall in the interval $\left(  0,y_{1}\right)  $, for a given $j\leq n$. This
event is shown in Table \ref{tab:decision}. Thus, to compute the probability
of interest, it is enough evaluate the cumulative distribution functions
relative to the event $B$ of the order statistic, given by
(\ref{eq:cumulative}). An efficient method for the computation of cumulative
distribution functions of order statistics from two populations was proposed
by Glueck et al. (2007). Here we need a slight generalization, involving the
event $B$, which requires a different proof.

\begin{theorem}
\label{thm:calc}Denote the index vector $\mathbf{i}=\left(  i_{0},i_{1},\ldots
i_{e+1}\right)  $ and the summation index set%
\begin{equation}
\mathcal{I=}\left\{  \mathbf{i:}%
\begin{array}
[c]{c}%
0=i_{0}\leq i_{1}\leq\cdots\leq i_{e}\leq i_{e+1}=m\text{, }\\
\text{and }i_{a}\geq n_{a}\text{ for all }1\leq a\leq k
\end{array}
\right\}  .
\end{equation}
Suppose that $F_{X_{i}}\left(  x\right)  =F\left(  x\right)  $, for all $1\leq
i\leq n,$ and $F_{X_{i}}\left(  x\right)  =G\left(  x\right)  $, for all
$n+1\leq i\leq m$. Then the cumulative distribution function relative to the
event $B$ (\ref{eq:cumulative}) is given by%
\begin{align}
&  F_{Y_{n_{1}},\ldots Y_{n_{e}};B}\left(  y_{1},\ldots,y_{e}\right)
\label{eq:perm}\\
&  \qquad=\sum_{\mathbf{i\in}\mathcal{I}}\sum_{\boldsymbol{\,\lambda}\,\,}%
\prod_{a=1}^{e+1}\frac{n!\left(  m-n\right)  !}{\lambda_{a}!\left(
i_{a}-i_{a-1}-\lambda_{a}\right)  !}\nonumber\\
&  \qquad\qquad\cdot\left[  F\left(  y_{a}\right)  -F\left(  y_{a-1}\right)
\right]  ^{\lambda_{a}}\left[  G\left(  y_{a}\right)  -G\left(  y_{a-1}%
\right)  \right]  ^{i_{a}-i_{a-1}-\lambda_{a}},\nonumber
\end{align}
where $y_{0}=0$, $y_{e+1}=1$, and $\boldsymbol{\lambda}=\left(  \lambda
_{1},\lambda_{2},\ldots,\lambda_{e+1}\right)  $ ranges over all integer
vectors such that $\lambda_{1}=j$ and%
\begin{equation}
\lambda_{1}+\lambda_{2}+\cdots+\lambda_{e+1}=n,\quad0\leq\lambda_{a}\leq
i_{a}-i_{a-1}.\label{eq:lambda-constr}%
\end{equation}

\end{theorem}

\begin{proof}
Denote by $A_{\mathbf{i,}\boldsymbol{\,\lambda}}$ the event that exactly
$i_{a}-i_{a-1}$ of the random variables $X_{i}$ fall in the interval
$(y_{a-1},y_{a})$, and exactly $\lambda_{a}$ of those are elements of $S_{1}$.
When $a=1$, $(y_{a-1},y_{a})=(y_{0},y_{1})=(0,y_{1}).$ If $B$ occurs,
$\lambda_{1}=j$. Then from the binomial theorem,
\begin{equation}
\Pr\left(  A_{\mathbf{i,}\boldsymbol{\,\lambda}}\right)  =\prod_{a=1}%
^{e+1}\frac{n!\left(  m-n\right)  !}{\lambda_{a}!\left(  i_{a}-i_{a-1}%
-\lambda_{a}\right)  !}\left[  F\left(  y_{a}\right)  -F\left(  y_{a-1}%
\right)  \right]  ^{\lambda_{a}}\left[  G\left(  y_{a}\right)  -G\left(
y_{a-1}\right)  \right]  ^{i_{a}-i_{a-1}-\lambda_{a}}.
\end{equation}

Since the events $A_{\mathbf{i,}\boldsymbol{\,\lambda}}$ for different
$\left(  \mathbf{i,}\boldsymbol{\,\lambda}\right)  $ are disjoint, the result follows.
\end{proof}

The only difference between Theorem \ref{thm:calc} and the result by Glueck et
al. (2007) is the added condition $\lambda_{1}=j$.

In the case of two random variables, we recover the same results as the direct
method in Section \ref{sec:example}. With $m=2$, $n=1$, $s=2$, $c_{1}=0$,
$d_{1}=b_{1}$, $c_{2}=b_{2}$, $d_{2}=1$, $S_{1}=\left\{  X_{1}\right\}  $,
$S_{2}=\left\{  X_{1}\right\}  $, $k_{1}=1$, $k_{2}=1$, $Y_{w_{1,1}%
}=Y_{w_{1,k_{1}}}=Y_{1}$, $Y_{w_{2,1}}=Y_{w_{2,k_{1}}}=Y_{2}$, using Theorem
\ref{theorem:1} and \ref{thm:calc} yields
\begin{equation}
\left(  E\cap B\right)  =\gamma,
\end{equation}
when $j=0$, and
\begin{equation}
\left(  E\cap B\right)  =\beta,
\end{equation}
when $j=1$.

In conclusion, for two sets of real valued, independent but not necessarily
identically distributed random variables, we have now given an expression for
the probability that arbitrary subsets of order statistics fall in disjoint,
ordered intervals, and that of the smallest statistics, a certain number come
from one set. \

\section{Concluding example}

The methods of this paper can be used to calculating the joint probability of
the number of rejections and the number of false rejection for the
Benjamini-Hochberg (1995) procedure. A rejection of a hypothesis for which the
null holds is a false rejection. Given an false discovery rate $\alpha
\in\left(  0,1\right)  $, hypotheses $H_{i}$ $i=1,\ldots,m$, $p$-values
$X_{i}$, and the corresponding order statistics for the $p$-values
$Y_{i}=X_{(e)}$ (the random variables $X_{i}$ sorted in nondecreasing order
$X_{(1)}\le X_{(2)}\le\cdots\le X_{(m)}$), the\ procedure produces a
nondecreasing sequence of numbers $b_{i}=i\alpha/m\in\left(  0,1\right)  $,
rejects the hypotheses $H_{(e)}$, $e=1,\ldots,k_{1}$, such that $k_{1}$ is the
largest number for which $y_{k_{1}}\leq b_{k_{1}}$, and accepts all others.
For $n\in\left\{  0,1,\ldots,m\right\}  $ assume that the null holds for
$H_{1},H_{2},\ldots,H_{n}$ and that the alternative holds for $H_{n+1}%
,H_{n+2},\ldots,H_{m}$. Let $S_{1}=\left\{  X_{1},X_{2},\ldots,X_{n}\right\}
$ be the set of p-values that correspond to the null hypotheses, and
$S_{2}=\left\{  X_{n+1},X_{n+2},\ldots,X_{m}\right\}  $ be the set of p-values
for which the alternative holds. Then $j$ is the number of null hypotheses
that are rejected, which is equal to the number of p-values corresponding to
null hypotheses that fall in the interval $[0,b_{k_{1}}]$.

Under the assumption that the p-values for which the alternative holds have
the same distribution, one can use the methods of this paper to find the joint
distribution of $j$ and $k_{1}$. For each value of $k_{1}$ and $m$, Glueck et
al. (2006a) pointed out that the rejection regions for the Benjamini and
Hochberg (1995) procedure can be decomposed into disjoint sets of events.
These events correspond to certain numbers of order statistics falling into
sets of intervals, defined by the numbers $b_{i}$. Details about the
decomposition of the rejection regions into these events are given in Glueck
et al. (2006a). The general case is too complicated to detail here. However, as
an example, we calculate the probabilities that with $m=2$ hypotheses, and
$n=1$ null hypotheses, the Benjamini and Hochberg (1995) procedure rejects
$k_{1}=1$ hypotheses, and that $j$, the number of false rejections, is either
$0$ or $1$.

Suppose we wish to test $m=2$ hypotheses. \ Specifically, we wish to test
hypotheses about the location of the sample mean. \ We plan to conduct a two
sided test. \ We assume that we have two large populations, with known
variances (both $\sigma^{2}$), and that the variables of interest, say
$\epsilon_{1}$ and $\epsilon_{2}$, are normally distributed, so that
\ $\epsilon_{1}\sim N\left(  \mu_{1},\sigma^{2}\right)  $ and $\epsilon
_{2}\sim N\left(  \mu_{2},\sigma^{2}\right)  $. \ We wish to test two
hypotheses $H_{1}:\mu_{1}=\mu_{_{0}}$, and $H_{2}:\mu_{2}=\mu_{_{0}}$, with
the alternative hypothesis for both populations the same, so $H_{A}:\mu
=\mu_{A}$. We sample $N_{i}$ random variables from each population, say
$\epsilon_{i1},\epsilon_{i2},\ldots\epsilon_{iN_{i}}$. \ For convenience, we
will assume that the random sample is of the same size for each hypothesis
test, so $N_{1}=N_{2}=N$.

With$\,$
\begin{equation}
\bar\epsilon_{i}=N^{-1}\sum_{\delta=1}^{N}\epsilon_{i\delta}\text{,}%
\end{equation}
the test statistics are given by
\begin{equation}
Z_{i}=\left(  \frac{\sigma}{\sqrt{N}}\right)  ^{-1}\left(  \bar\epsilon
_{i}-\mu_{0}\right)  ,
\end{equation}
and the two sided p-values are (Rosner, p. 244, 2006)
\begin{equation}
X_{i}=
\begin{cases}
2\Phi\left(  Z_{i}\right)  & Z_{i}\leq0\\
2\left[  1-\Phi\left(  Z_{i}\right)  \right]  & Z_{i}>0
\end{cases}
\text{ },
\end{equation}
where $\Phi$ is the cumulative distribution function of \ the standard normal
(mean = 0 and variance \ = 1). \ Let $\phi$ be the probability density
function of the standard normal. \newline Suppose that in truth, we have
$\epsilon_{1}\sim N\left(  \mu_{0},\sigma^{2}\right)  $, so that the null
holds for $H_{1}$, and $\epsilon_{2}\sim N\left(  \mu_{A},\sigma^{2}\right)
$, so that alternative holds for $H_{2}$. \ \ Define $S_{1}=\left\{
X_{1}\right\}  $, and $S_{2}=\left\{  X_{2}\right\}  $. \ Then the number of
p-value for which the null holds, $n=1$. For $H_{1}$, the hypotheses for which
the null holds, the p-value has a uniform distribution on the interval
$\left[  0,1\right]  $, so for $x_{1}\in\left[  0,1\right]  $,
\begin{equation}
F_{X_{1}}\left(  x_{1}\right)  =x_{1}.
\end{equation}

For $H_{2}$, the alternative holds. \ When we conduct the hypothesis test, we
are unaware of the truth. We always calculate the p-value under \ the null.
\ However, since the alternative actually holds,
\begin{equation}
\label{cdf}\begin{aligned}[t] \Pr\left[Z_2\leq z_{_2}\right]=&\Pr\left[\frac{\bar \epsilon_i-\mu_0}{\frac{\sigma}{\sqrt{N}}}\leq z_{_2}\right]\\ =&\Pr\left[\frac{\epsilon_i-\mu_A}{\frac{\sigma}{\sqrt{N}}}\leq z_{_2}+\frac{\mu_0-\mu_A}{\frac{\sigma}{\sqrt{N}}}\right]\\ =&\Phi\left[z_{_2}+\frac{\mu_0-\mu_A}{\frac{\sigma}{\sqrt{N}}}\right]. \end{aligned}
\end{equation}
Finally,
\begin{equation}
\begin{aligned}[t] F_{X_2}\left(x_2\right)=&\Pr\left(X_2<x_2\right)\\ =&\Pr\left(\left\{X_2<x_2\right\}\cap\left\{Z_2\leq 0\right\}\right)+\Pr\left(\left\{X_2<x_2\right\}\cap\left\{Z_2>0\right\}\right)\\ =&\Pr\left(\left\{2\Phi\left(Z_2\right)<x_2\right\}\right)+\Pr\left(\left\{2\left[1-\Phi\left(Z_2\right)\right]<x_2\right\}\right)\\ =&\Pr\left(\left\{Z_2\leq\Phi^{-1}\left(x_2/2\right)\right\}\right)+1-\Pr\left(\left\{Z_2\leq\Phi^{-1}\left(1-x_2/2\right)\right\}\right)\\ =&\Phi\left[\Phi^{-1}\left(x_2/2\right)+\frac{\mu_0-\mu_A}{\frac{\sigma}{\sqrt{N}}}\right]+1-\Phi\left[\Phi^{-1}\left(1-x_2/2\right)+\frac{\mu_0-\mu_A}{\frac{\sigma}{\sqrt{N}}}\right], \end{aligned}
\end{equation}
where the last step follows by substitution from Equation \ref{cdf}.

Now, as a specific example, we fix $\mu_{0}=0$, $\mu_{A}=1$, $\sigma^{2}=1$
$\alpha=.05$. We wish to calculate the probability that $k_{1}=1$, and that
$j=0$ or $j=1$. With $c_{1}=0$, $d_{1}=\alpha/2$, $c_{2}=\alpha$, $d_{2}=1$.
This is the probability that of the two hypotheses, we reject exactly one, and
it is $H_{1}$, the hypothesis for which the null holds. When $j=0$, the
rejection we make is of the hypothesis for which the alternative holds, and
when $j=1$, the rejection we make is of the null hypothesis, a false rejection.

We calculated the probability using our methodology, and by a simulation using
a sample of 100,000 variables. Recall that $k_{1}$ is the number of order
statistics that are less than $b_{1}$, and $j$ are the number in Set 1, and
less than $b_{1}$. The results are shown in Table \ref{back}.

\begin{table}[ptb]
\begin{center}%
\begin{tabular}
[t]{|l|l|l|l|l|}\hline
$k_{1}$ & $j$ & Theory & Simulation & Difference\\\hline
1 & 0 & .472982 & .47388 & .000898\\\hline
1 & 1 & .00978051 & .0095 & .00028051\\\hline
\end{tabular}
\end{center}
\caption{Comparison of Simulation and Theory. \ Recall that $k_{1}$ is the
number of hypotheses that were rejected, and $j$ is the number of null
hypotheses that were rejected. We had two hypotheses, and one null
hypothesis.}%
\label{back}%
\end{table}

Notice that the simulation differs from the theory only in the fourth decimal
place. The theory is exact. Software that implements this method in
Mathematica is available from the authors upon request.

\section*{References}
\hspace*{\parindent}%
Bapat, R. B. and Beg, M. I. (1989). ``Order Statistics for non-identically
distributed variables and permanents,'' \textit{Sankhya}, Ser. A., 51, 79-93.

Benjamini, Y., and Hochberg, Y. (1995). Controlling the False Discovery Rate:
A Practical and Powerful Approach to Multiple Testing. \textit{J. R. Stat.
Soc. Ser. B Stat. Methodol.} 57 289-300.

David, H. A. (1981). \textit{Order Statistics}, (2nd ed.). New York: Wiley.

Glueck, Deborah H., Muller, Keith E., Karimpour-Fard, Anis, Hunter, Lawrence.
(2006a) (in review), \ Expected Power for the False Discover Rate with Independence.

Glueck, D. H., Karimpour-Fard, A., \ Mandel, \ J. and Muller, K.E. (2006b)
(in review), \ On the probability that order statistics fall in intervals.

Glueck, D. H., Karimpour-Fard, A., \ Mandel, \ J. , Hunter, \ L. and Muller,
K.E. (2007) (in review), \ Fast computation by block permanents of cumulative distribution
functions of order statistics from several populations. \emph{arXiv:0705.3851}

Rosner B. (2006). \textit{Fundamentals of Biostatistics} (6th edition). \ New
York: Brooks-Cole.

Ross, S. (1984). \textit{A First Course in Probability: Second Edition}. New
York: Macmillan Publishing Company.

\end{document}